\RequirePackage{lineno}
\documentclass[11pt,jkps,preprint,fleqn,showpacs,showkeys]{revtex4}
\usepackage[pdftex]{graphicx}
\usepackage{amssymb}
\usepackage{amsmath}
\usepackage{bm}
\usepackage[fleqn]{nccmath}
\usepackage{caption}
\usepackage{tabularx}
\usepackage{esvect}
\MakeRobust{\vv}
\usepackage{subfigure}
\usepackage{url}
\begin{document}
\setcounter{page}{0}
\title[]{A fabrication of low power distribution system for muon $g - 2$/EDM experiment at J-PARC}
\author{E. \surname{WON}}
\author{WOODO \surname{LEE}}
\email{woodolee@korea.ac.kr}
\thanks{Tel: +82-10-7517-2711}
\affiliation{Department of Physics, Korea University, Seoul
06130, Korea}

\begin{abstract}

The J-PARC muon $g - 2$/EDM experiment aims to measure the muon magnetic moment anomaly $a_{\mu} = (g -2) / 2 ~ $ and the muon electric dipole moment (EDM) $d_{\mu}$.
The target sensitivity for $a_{\mu}$ is a statistical uncertainty of $450 \times 10^{-9}$, and for $d_{\mu}$ it is $1.5 \times 10^{-21}$ $e \cdot \rm{cm}$. 
The readout electronics require a DC to DC converter to provide 1.5 V and 3.3 V electric voltage within a specific environment. 
To capture the positron decay from the muon, the detector components are placed in the muon storage area where 3 T magnetic field is applied in a vacuum. 
The extra magnetic field from the converter should be less than 30 $\mu$T to reach the aforementioned sensitivity on $a_{\mu}$.
In addition, the electric field produced by the converter has to be as small as 1 V/m to also reach the target sensitivity on $d_{\mu}$.
For this purpose, we develop a step-down converter.
We discuss the development process and the performance of our DC to DC converter for the J-PARC muon $g - 2$/EDM experiment.
\end{abstract}


\maketitle

\section{INTRODUCTION}

The J-PARC muon $g - 2$/EDM experiment \cite{10.1093/ptep/ptz030} aims to measure the muon magnetic moment anomaly $a_{\mu} = (g -2 ) / 2 ~ $ and the muon electric dipole moment (EDM) $d_{\mu}$, which requires a specific low direct current (DC) power distribution system with a minimum stray field generation of as low as 30 $\mu$T and 1 V/m in the muon orbit system. 
It also has to provide appropriate electric powers to the readout system under a strong 3 T magnetic field. 
Here we develop a DC to DC converter that meets all the requirements from the experiment.
The muon $g - 2$/EDM experiment detects positrons with silicon detectors, and the electric readout system requires DC power distribution for its operation. 
The DC to DC converter is designed to generate various DC outputs under a strong static magnetic field. 
Our design of the DC to DC converter utilizes buck converter \cite{DCDC} as well, which is commonly used in high energy physics \cite{CMSDCDC, JPARCDCDC}.
However, our design is carefully optimized to mitigate the stray magnetic field for the experiment, and we discuss the design details in the following sections.

\section{Requirements and design}

There are several requirements in the design of the DC to DC converter for the J-PARC muon $g - 2$/EDM experiment.
First, appropriate electric power must be delivered to the readout electronics.
Second, in the design of the converter, the space limitation due to the geometry of the detector system, including the readout electronics, must be carefully considered.
These requirements allow us to use only limited board geometry and electric components.
They will be discussed in the following sections.

\subsection{Electric power requirements}

The DC to DC converter has to provide electric powers to four Application Specific Integrated Circuit chips (ASICs), one Field-Programmable Gate Array (FPGA), and one optical transceiver for the operation of the readout electrics system of the experiment.
Each receives different voltages and current values for the operation, and they are summarized in Table \ref{table:requirements}.
Both an optical transceiver and a FPGA receive 1.0 A and 3.3 V.
However, for ASICs, they receive 0.9 A and 1.5 V.
The DC to DC converter is designed to provide stable electric power of $8$ W in total, to the readout system. 

\begin{table}
\begin{tabular}{ c c c c }
 \hline
 \multicolumn{3}{c}{Electric power consumption list} \\
 \hline
 Component  & Current (A) & Voltage (V)\\
 \hline
 Optical transceiver  & 1.0    &  3.3\\
 ASICs    & 0.9    &  1.5    \\
 FPGA    & 1.0    &  3.3    \\
 \hline
\end{tabular}
\caption{
List of electric components in the readout system that receives electric powers from the DC to DC converter.
Both an optical transceiver and an FPGA receive 1.0 A and 3.3 V.
However, ASICs receive 0.9 A and 1.5 V.
}
\label{table:requirements}
\end{table}

\subsection{Space requirements}

\begin{figure}[h!]
\includegraphics[width=\textwidth]{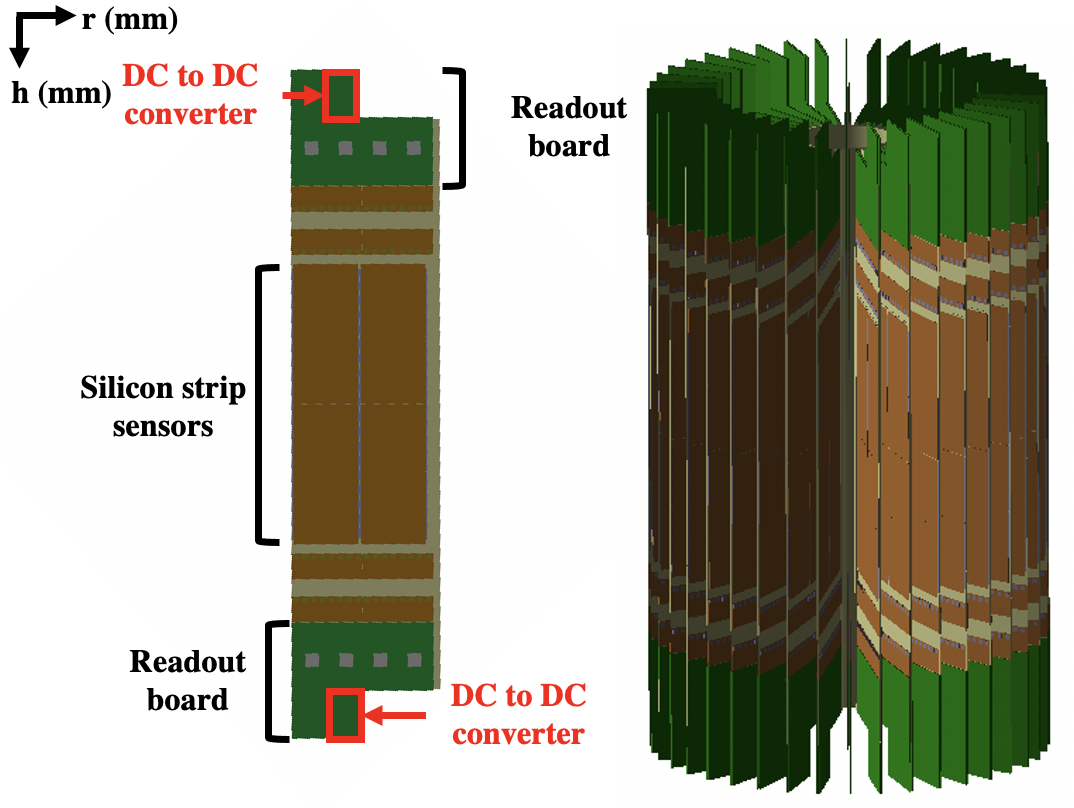} 
\caption{
The positron detector consists of in total 40 modules.
One module contains four arrays of silicon strip sensors. 
The distance from the center of the DC to DC converter to the vane center is $435$ mm.
}
\label{fig:detectorStruct}
\end{figure}

The positron detector modules in the $g - 2$/EDM experiment are placed in the cylindrical storage under a uniform magnetic field. 
In total, 40 detector modules will be mounted in radial direction like vanes, as shown in Fig. \ref{fig:detectorStruct}.
Each vane has the following structure. 
The central part consists of the strip silicon sensor, and two readout systems are placed at the top and bottom position of silicon strip sensor arrays.
The DC to DC converter will be mounted at the top and bottom of the readout system. 
Due to this geometrical configuration, the height of the DC to DC converter should be smaller than the gap between each vane.
Along with this constraint, the size of the converter should be as small as 37 $\times$ 180 $\times$ 150 $\rm{mm^{3}}$ for the height, the width, and the length, respectively.

\subsection{The board design}

\begin{figure}[h!]
    \centering
    \includegraphics[width=0.8\textwidth]{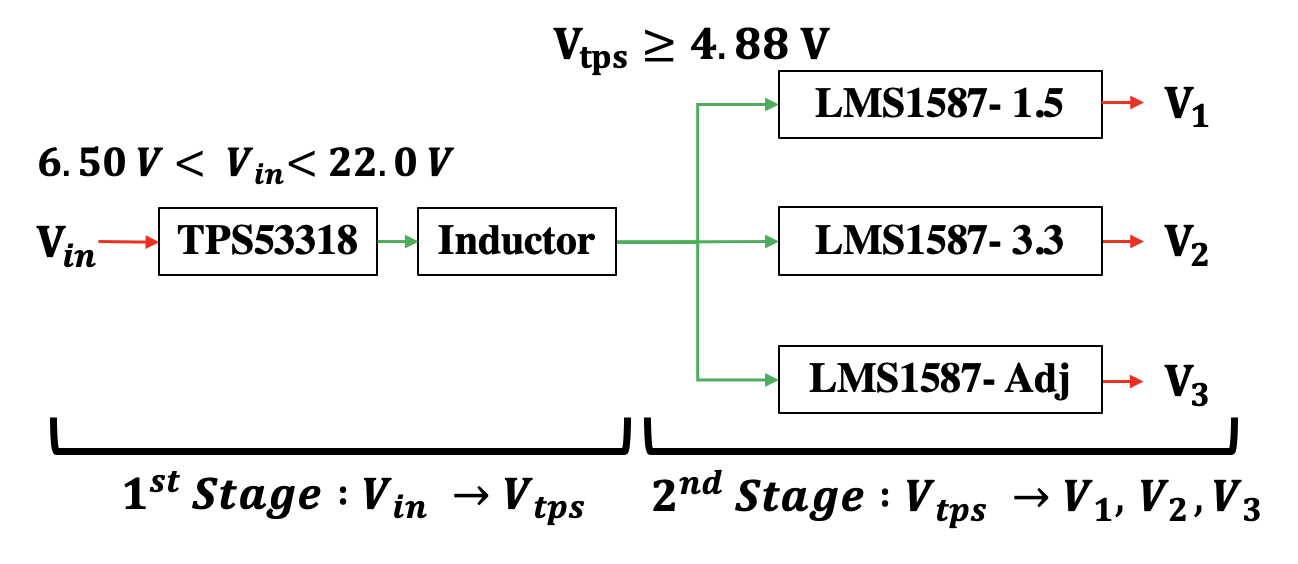}
    \caption{
    The design diagram for the low power distribution is shown. 
    We design the low-power distribution system with two distinct stages.
    The TPS53318 is for the first stage and three different linear regulators are for the second stage, and outputs are to be delivered to the readout electronics.
    }
    \label{fig:diagram}
    \end{figure}

Our board has two main stages according to its electric functionality and are shown in Fig. \ref{fig:diagram}.
The $V_{in}$ is defined as the input voltage for the DC to DC converter at the first stage in this figure.
The first stage is composed of a switching type DC to DC converter.
Due to the fact that high current is applied to the first stage, we choose this switching type of DC to DC converter to minimize the electric power loss.
We choose to use a product from Texas Instrument of the model TPS53318 \cite{TPS53319} in order to achieve high efficiency in a high current environment.
The TPS53318 chip drops down $V_{in}$ to $V_{tps}$, and $V_{tps}$ is defined as the output voltage of the TPS53318.
Note that switching type converters require an associated inductor, and this forces us to design an inductor carefully.
We discuss it in the next section.
The value of $V_{tps}$ is set to 4.88 V, which is the minimum voltage that the linear regulators can receive at the second stage.
At the second stage, the current is divided into three different channels, and the output of each channel is directly connected to the electric devices in the readout electronics.
We choose three different linear type regulators for suppressing electric noise.
The outputs of the regulators are 1.5 V, 3.3 V, and variable DC voltage, which are referred to as $V_{1}$, $V_{2}$, and $V_{3}$.
The specifications of the TPS53318 chip and three linear regulators are summarized in Table \ref{table:IC_parts}.

\begin{figure}[h!]
\includegraphics[width=\textwidth]{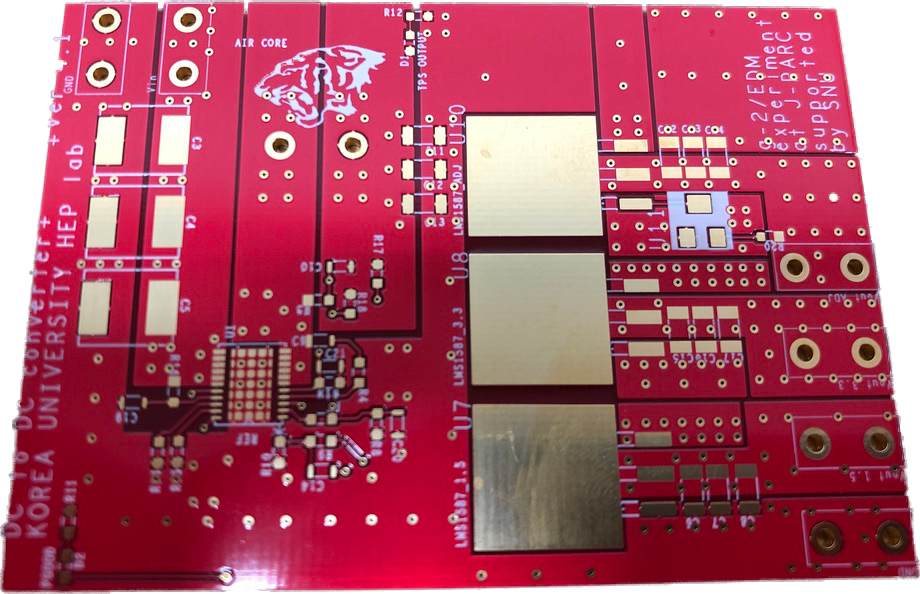}
\caption{
The PCB for the DC to DC converter is shown. 
The board consists of 4 layers, the thickness of the board is $1.5 ~ \mathrm{mm}$. 
There are two middle layers between the top and bottom layers. The middle layers are for input, output power, and ground. 
}
\label{fig:boards}
\end{figure}

The size of Printed Circuit Board (PCB) and its number of layers for the DC to DC converter are designed to satisfy the space and electric requirements as discussed above.
Figure \ref{fig:boards} shows the fabricated PCB. 
The area of the board is determined to be 180 $\times$ 150 $\rm{mm^{2}}$, and the board has four layers for routing signals and grounds to all the components.

\begin{table}[h!]
\begin{tabular}{  c c c  }
 \hline
 \multicolumn{3}{ c }{Main IC chips for DC to DC converter} \\
 \hline
  IC & $V_{in}$ (V) & $V_{out}$ (V)  \\
 \hline
 TPS53318     & 6.50 $\sim$ 22.00 & 0.6 $\sim$ 5.5  \\
 LMS1587-3.3   & 4.75 $\sim$ 7.00 & 3.3        \\
 LMS1587-1.5   & 3.00 $\sim$ 7.00 & 1.5         \\
 LMS1587-Adj   &  2.75 $\sim$ 7.00 &  1.5 $\sim$ 5.8 \\
 \hline
\end{tabular}
\caption{
The specifications of main Integrated Circuit (IC) chips for DC to DC converter are shown.
The TPS53318 is for the first stage, and three different linear regulators (LMS1587 series \cite{LMS1585}) are for the second stage, and outputs are to be delivered to the readout electronics.
}
\label{table:IC_parts}
\end{table}

\subsection{The inductor design} 

Our DC to DC converter is required not to introduce the magnitude of the stray magnetic field of more than 1 parts per million (ppm) \cite{10.1093/ptep/ptz030} of the 3 T static magnetic field.
Because of this, the commonly used ferrite core inductor cannot be used since the ferrite core itself is magnetized, resulting in the production of the stray magnetic field. 
The second problem is that its inductance is changed by the external magnetic field. 
This may results in the faulty operation of the DC to DC converter. 
An air-toroidal inductor (ATI) with a copper coil is fabricated instead of the ferrite core to avoid these problems.
The suitable physical size of ATI, which is limited by the gap between two adjacent vanes, is discussed in this section.
The required inductance ($L$) of ATI for given maximum output current ($I^{MAX}_{out}$), switching frequency ($f_{sw}$) of the converter, maximum input voltage ($V_{in}^{MAX}$)
from our DC to DC voltage is provided by the company \cite{TPS53319} following formula
\begin{ceqn}
\begin{equation}
    L  = \frac{3}{I^{MAX}_{out} \cdot f_{sw}} \cdot \frac{(V_{in}^{MAX} - V_{tps})\cdot V_{tps} }{V_{in}^{MAX}}.
\end{equation}
\label{equation:eq_L}
\end{ceqn}

\begin{figure}[h!]
\includegraphics[width=\textwidth]{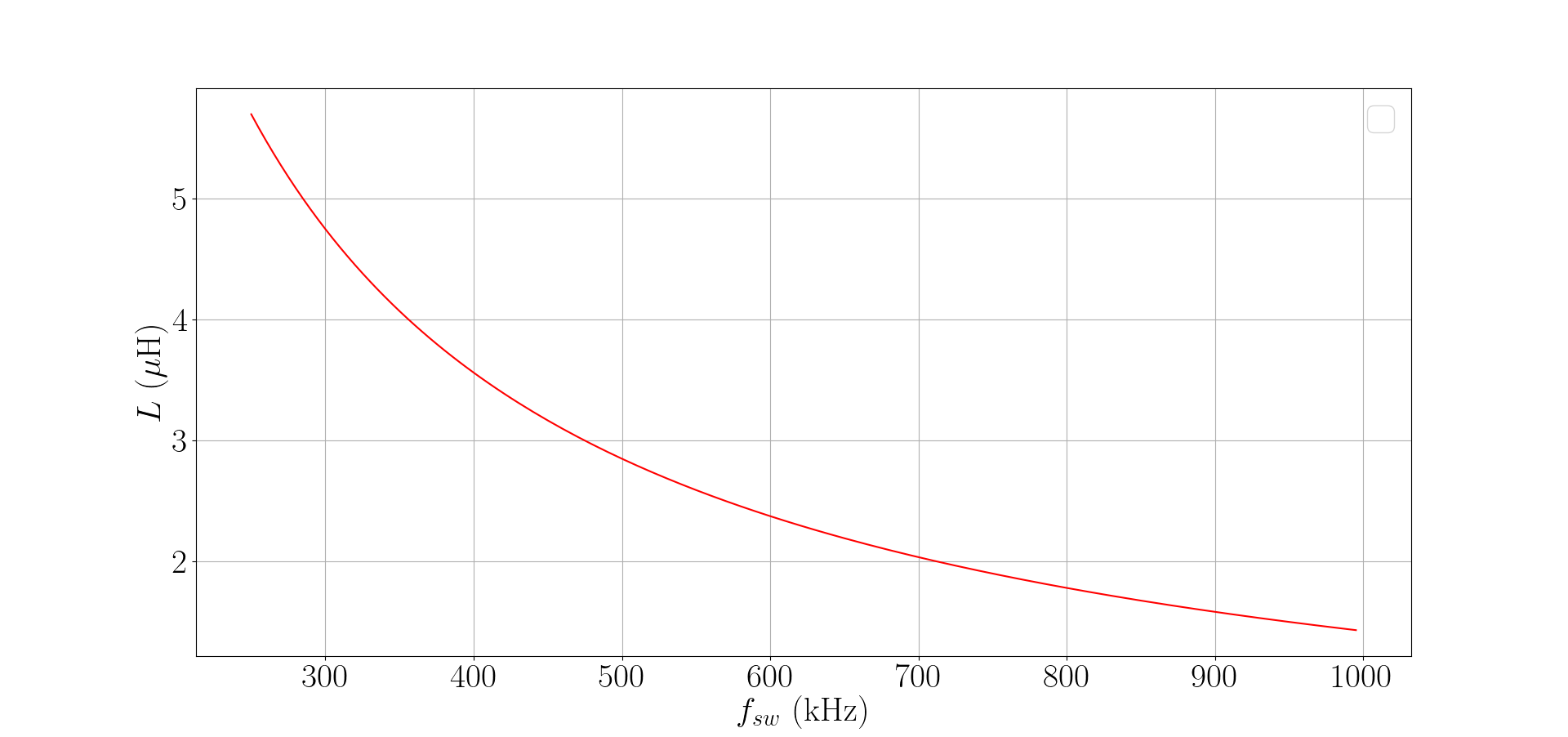}
\caption{
The expected inductance $L$ as a function of $f_{sw}$ is shown.
Note that the recommended range of $f_{sw}$ according to the company is from 250 to 970 kHz.
}
\label{fig:fVsL}
\end{figure}

Figure \ref{fig:fVsL} shows the relationship between $L$ and $f_{sw}$ when we assume other values to be for $V_{tps}$ = 4.88 V, $V_{in}^{MAX} = 22 ~ \mathrm{V}$, and $I^{MAX}_{out} = 8 ~ \mathrm{A}$ (they are allowed maximum values of given chip).
Since the minimum $f_{sw}$ of our chip is 250 kHz, and the power loss from the switching is minimized with the lowest $f_{sw}$, our target $f_{sw}$ is set to 250 kHz and this results in our target value of $L$ as $5.70 ~ \mu \rm{H}$ according to Fig. \ref{fig:fVsL}.
 
\begin{figure}[h!]
\centering
\includegraphics[width=\textwidth]{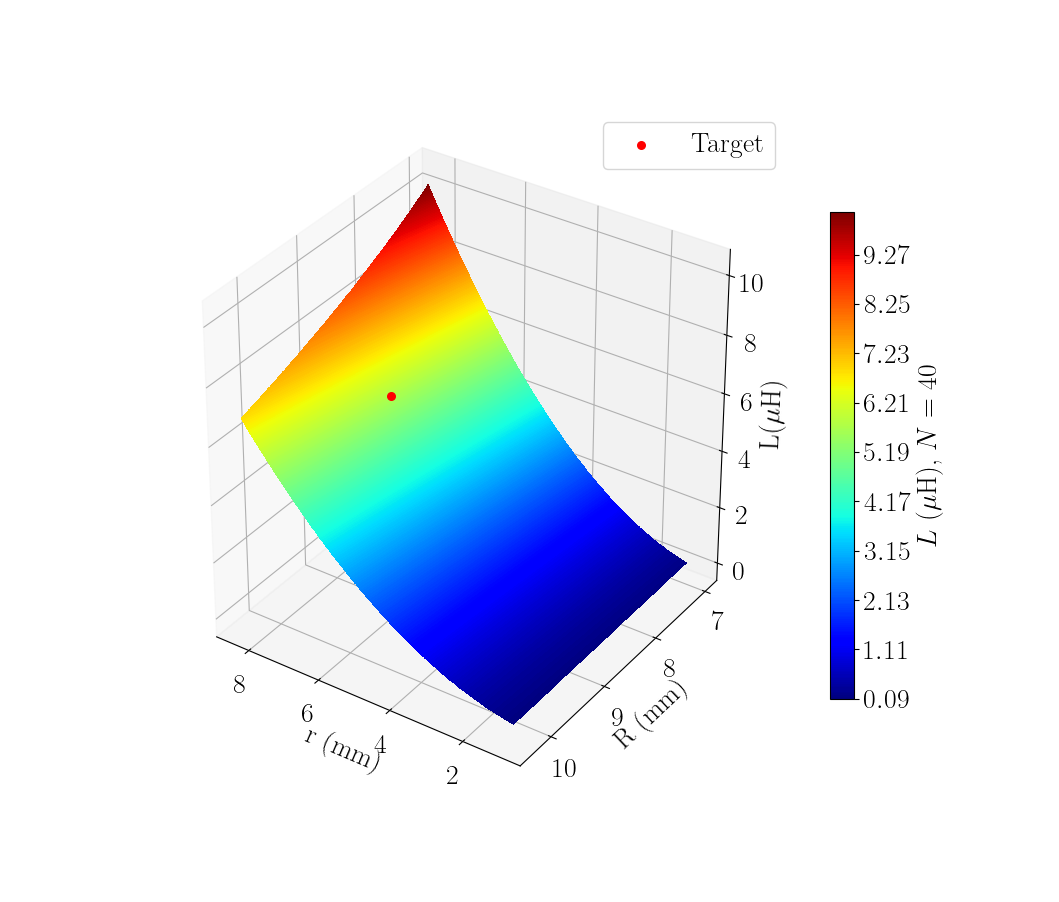}
\caption{
A relationship among $r$, $R$, and $L$ with $N$ = 40 is shown.
The target point with $L$ = 5.70 $\mu \mathrm{H}$ is shown as a circle.
}
\label{fig:geoCoil}
\end{figure}

There are three parameters that determine the inductance value of the ATI ($L_{ATI}$). 
They include the outer radius ($R$) from the center of the coil to the center of the cross-section and the radius of the cross-section ($r$) itself.
The third parameter is the number of turns ($N$) of the coil.
Considering the height limitation between the DC to DC converter to the next vane, we choose $r$ to be 7 mm.
Considering the area of the converter, we choose $R$ to be 8.6 mm.
This gives us $L_{ATI} = 5.70 ~ \mu \rm{H}$ with $N$ = 40.
Figure \ref{fig:geoCoil} shows the relationship among $r$, $R$, and $L$ when $N$ = 40.
There may be other sets of solutions for the target inductance, but it is not important for similar values of $r$ and $R$, so our default values for $L_{ATI}$ are set to be these values.
 
\section{Simulation}

\subsection{Electric circuit simulation}

\begin{figure}[h!]
\centering
    \includegraphics[width=\textwidth]{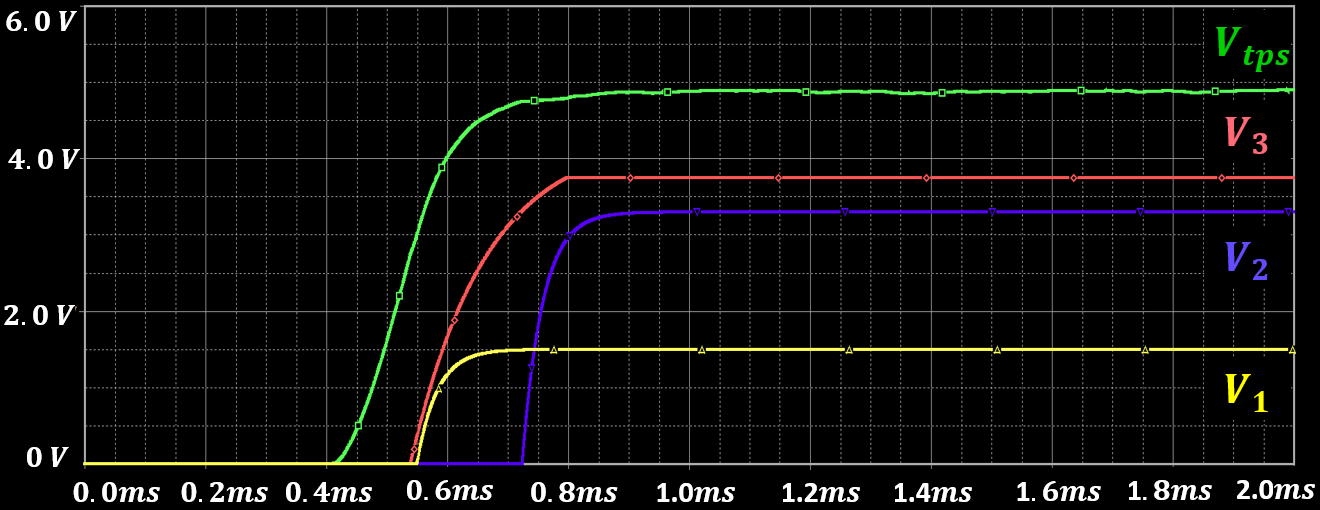}
    \caption{
    The electric characteristics of the DC to DC converter are quantified by simulation.
    $V_{tps}$ is the output voltage from the TPS53318, and $V_{1}$, $V_{2}$ is fixed-output voltage from the linear regulators as 1.5 V and 3.3 V, respectively.
    $V_{2}$ means variable-output voltage, which tunes to 3.8 V.
   }
    \label{fig:pspice}
\end{figure}

The electric characteristic of the DC to DC converter is studied by the PSPICE simulation  \cite{allegro}.
Our simulation estimates the time structure at a transient-state and the voltage fluctuation.
This simulation also aims at studying the stability of $V_{tps}$, $V_{1}$, $V_{2}$, and $V_{3}$ when $L_{ATI} = 5.70 ~ \mu \mathrm{H}$ and at a steady-state.
The DC to DC converter enters to transient-state at $0.42 ~ \rm{ms}$ from the power turns on, and after that, the DC to DC converter becomes a steady-state from $0.82 ~ \rm{ms}$ as shown in Fig. \ref{fig:pspice}.
The three outputs from linear regulators, $V_{1}$, $V_{2}$, and $V_{3}$ become a steady-state after $V_{tps}$ becomes stable.  
The $V_{1}$ goes into a steady-state first because it requires the minimum input voltage among the linear regulators.
The $V_{2}$ and $V_{3}$ reach at the steady-state after $V_{1}$ reaches to the steady state.
We compute the fluctuation value of a given voltage to be 1 $\sigma$ of the histogram made from the voltage itself.
From this, we compute fluctuation level, defined as normalized fluctuation by the mean voltage value, and for each output voltages at steady-states are shown in Table \ref{table:steady}, according to our simulation results.
The normalized fluctuation levels are reduced from 3.6$\times 10^{-3}$ to the order of $10^{-5}$, due to linear regulators.
Note that the maximum voltage fluctuations required by FPGA, ASICs, and optical transceiver in the readout system is 5 $\%$, and therefore the output of the DC to DC converter are much smaller than requirements, at least in the simulation stage.

\begin{center}
\begin{table}[h!]
\centering
\begin{tabular}{  c c c c  }
\hline
 Channels & Mean (V) & $\sigma$ (V) & Fluctuation level  \\
 \hline
    $V_{tps}$ & 4.95 & 1.80 $\cdot 10^{-2}$ & 3.6 $\cdot 10^{-3}$  \\
    $V_{1}$ & 1.49 &  2.18 $\cdot 10^{-5}$ & 1.4 $\cdot 10^{-5}$ \\
    $V_{2}$ & 3.29 &  2.24 $\cdot10^{-5}$ & 6.8 $\cdot 10^{-5}$ \\
    $V_{3}$ & 3.78 &  2.22 $\cdot10^{-5}$ & 5.8 $\cdot 10^{-5}$\\
 \hline
\end{tabular}
\caption{
The mean values, their fluctuations, and their normalized fluctuations levels are shown for different voltage outputs.
}
\label{table:steady}
\end{table}
\end{center}

\subsection{Simulation of the stray magnetic field}

\begin{figure*}[!t]
\centering
\subfigure[]{
\includegraphics[width=.45\columnwidth]{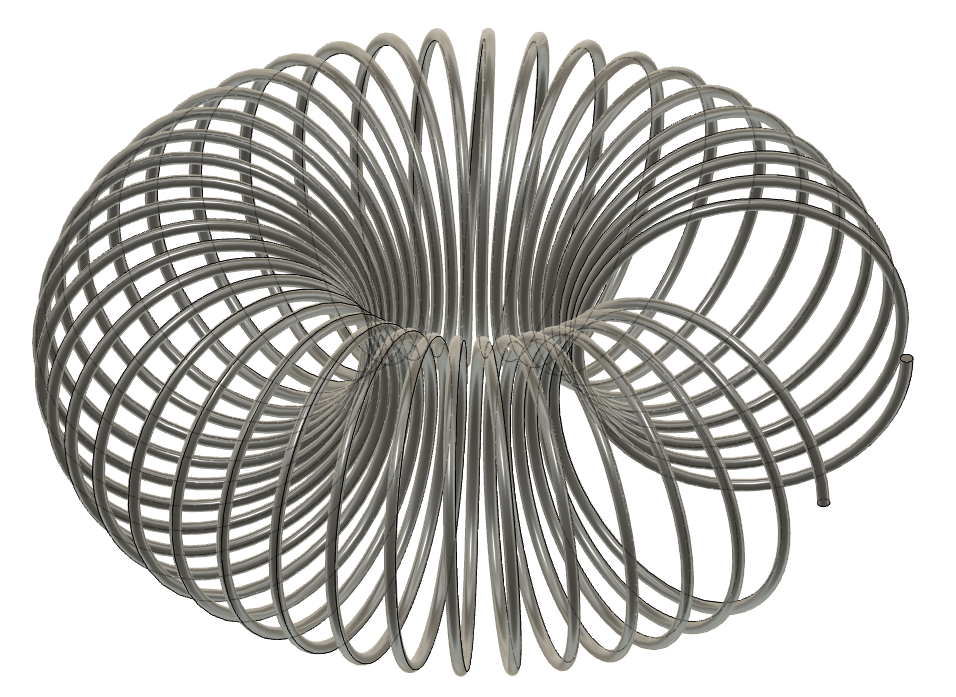}
}
\subfigure[]{
\includegraphics[width=.45\columnwidth]{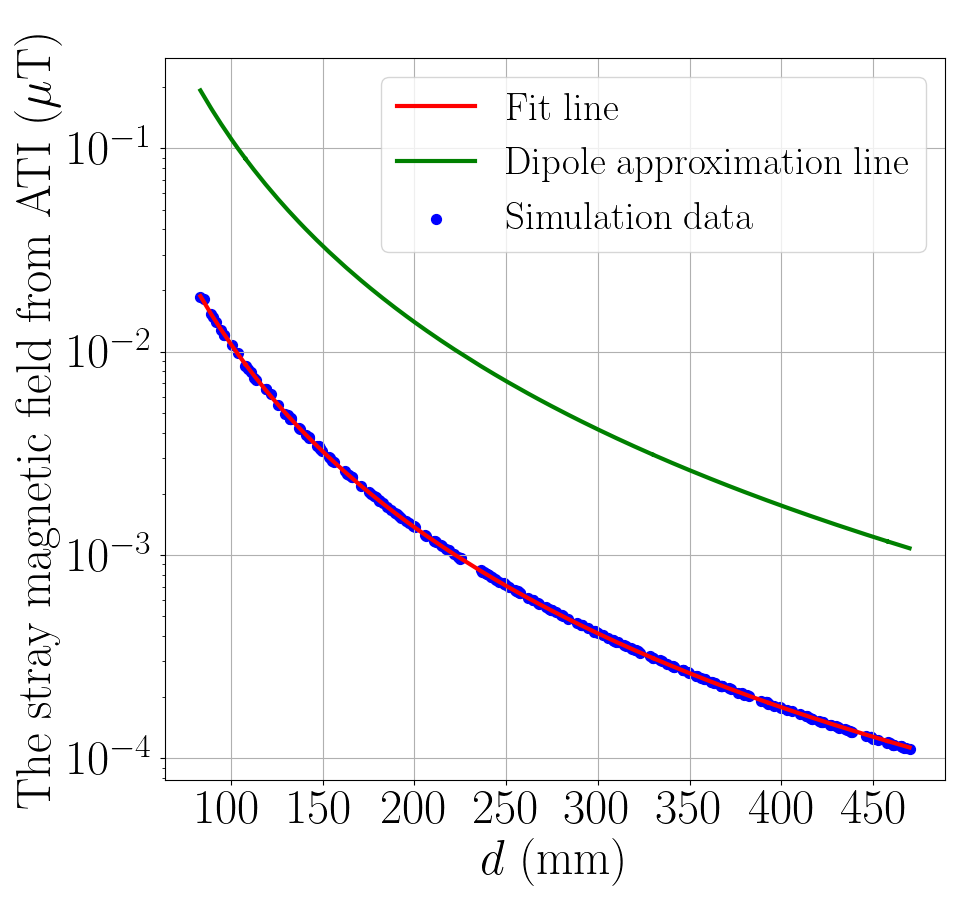}
}
\caption{
(a) The geometry of the ATI for the simulation is shown.
The ATI is drawn with Fusion 360 \cite{fusion360}, and $L_{ATI} = 5.70 ~ \mu$H with $r$ = 7 mm, $R$ = 8.6 mm, and $N$ = 40.
(b) A simulation result (closed circles) and a fit to the result (bottom curve) are shown.
A calculated inductance value from a dipole approximation is also shown (top curve).
The magnitude of the stray magnetic field is obtained by COMSOL, which is represented with blue dots.
A red line means a fitting line for the results.
}
\label{fig:comsol_B}
\end{figure*}

As we discussed above, it is extremely important to minimize the stray magnetic field from the DC to DC converter, and the dominant source of the stray magnetic field is likely from the ATI. 
Here, the magnitude of the stray magnetic field from the ATI is computed by our simulation.
First, we assume the maximum current of the ATI is 5 A when four ASICs and one FPGA are active.
The stray magnetic field from the ATI is simulated by COMSOL \cite{comsol}.
The geometry that is used in our simulation is shown in Fig. \ref{fig:comsol_B} (a).
Here, we use $L_{ATI} = 5.70 ~ \mu$H with $r$ = 7 mm, $R$ = 8.6 mm, and $N$ = 40.
The result of our simulation is shown in Fig. \ref{fig:comsol_B} (b).
A simulation result (closed circles) and a fit to the result (bottom curve) are shown.
A calculated inductance value from the dipole approximation of out ATI is also shown (top curve).
This dipole approximation of the ATI can be considered as an upper limit of the stray magnetic field.
The magnitude is decreased $\propto$ $d^{-3}$, where $d$ is the distance from the center of ATI to the point of interest.
Note that the dependence of $d^{-3}$ is from the dipole radiation of the electromagnetic field.
Considering the constant 3 T magnetic field is applied in the muon storage path, the effect from the stray magnetic field due to the ATI is estimated to be 0.03 ppb at $d$ = 435 mm where the muon storage area is located.
We attempt to measure the magnetic field from the ATI directly using LakeShore 425 Gauss meter \cite{Lakeshore}.
The non-zero stray magnetic field is seen at the very front of the ATI.
However, it is impossible to measure at a distance where the muon storage path is located since the magnitude is too small. 
However, at $d$ = 6 mm, the measured magnetic field is found to be 256 $\mu$T. 
From the simulation, the magnetic field at $d$ = 6 mm is about 267 $\mu$T which is somewhat larger than the measured value.
This is probably from the fact that the actual current applied is smaller than 5 A for the measurement, or discrepancy between real coil geometry and the geometry implemented in COMSOL.

\section{Fabrication and performance}

\begin{figure}[h!]
\centering
  \includegraphics[width=0.5\textwidth]{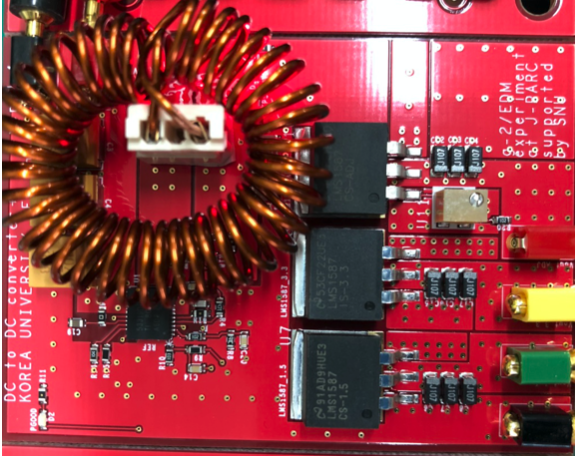}
    \caption{
     The fabricated DC to DC converter is shown. 
     The size of the DC to DC converter is $70 \times 50 ~ \rm{mm^{2}}$ and the height is $15 ~ \rm{mm}$.
    }\label{fig:wholeBoards}
\end{figure}

Our DC to DC converter with previously discussed ATI is fabricated, and Fig. \ref{fig:wholeBoards} shows a view of the converter.
In the figure, the ATI, TPS53318 (below the ATI), and three linear regulators can be seen.
There are three output channels to the readout electronics.
Two channels produce two values of voltages (3.3 V and 1.5 V), respectively.
As a spare, the third channel is prepared for a variable output voltage.
In what follows, we study the performance of the converter with the dummy loads to realize the environment when ASICs and FPGA are connected to the DC to DC converter on $V_{1}$ and $V_{2}$.
Two resistors with the resistance value of 1.1 $\Omega$ are connected to 1.5 V output and 3.3 V output to generate 4.3 A as a total sum of output currents.
We also study our converter for heat dissipation. 

\subsection{Voltage stability}

The DC to DC converter is tested to check the stability of output voltages and the efficiency of the converter. 
The output voltage should be stable at steady-state because the output from DC to DC converter will be directly connected to electronics in the readout system.
Unlike the simulation that we discussed previously, there is extra voltage fluctuation from the internal switching signal in the TPS53318, and it is not observed in the PSPICE simulation.
Therefore, for the real measurement, we defined the voltage fluctuation ($\delta V $) as 
\begin{ceqn}
\begin{equation}
\delta V (\%) =\frac{ V^{p-p} }{ V^{out}}  \cdot 100 , 
\end{equation}
\end{ceqn}
where $V^{out}$ is the output voltage of the DC to DC converter, and $V^{p-p}$ is the peak-to-peak value of the fluctuation in $V^{out}$.
We measure $\delta V $ when the input voltage is set to 22 V.
Our target value of $\delta V $ is set to be $\leq$ 5 \%.
For the channel with 1.5 V output, $\delta V $ is found to be 3.3 \%, and with 3.3 V output $\delta V $ is found to be 1.7 \%. 
Therefore, both channels satisfy our voltage fluctuation requirement.

\subsection{Measurement of efficiency}

\begin{figure}[h!]
\centering
   \includegraphics[width=\textwidth]{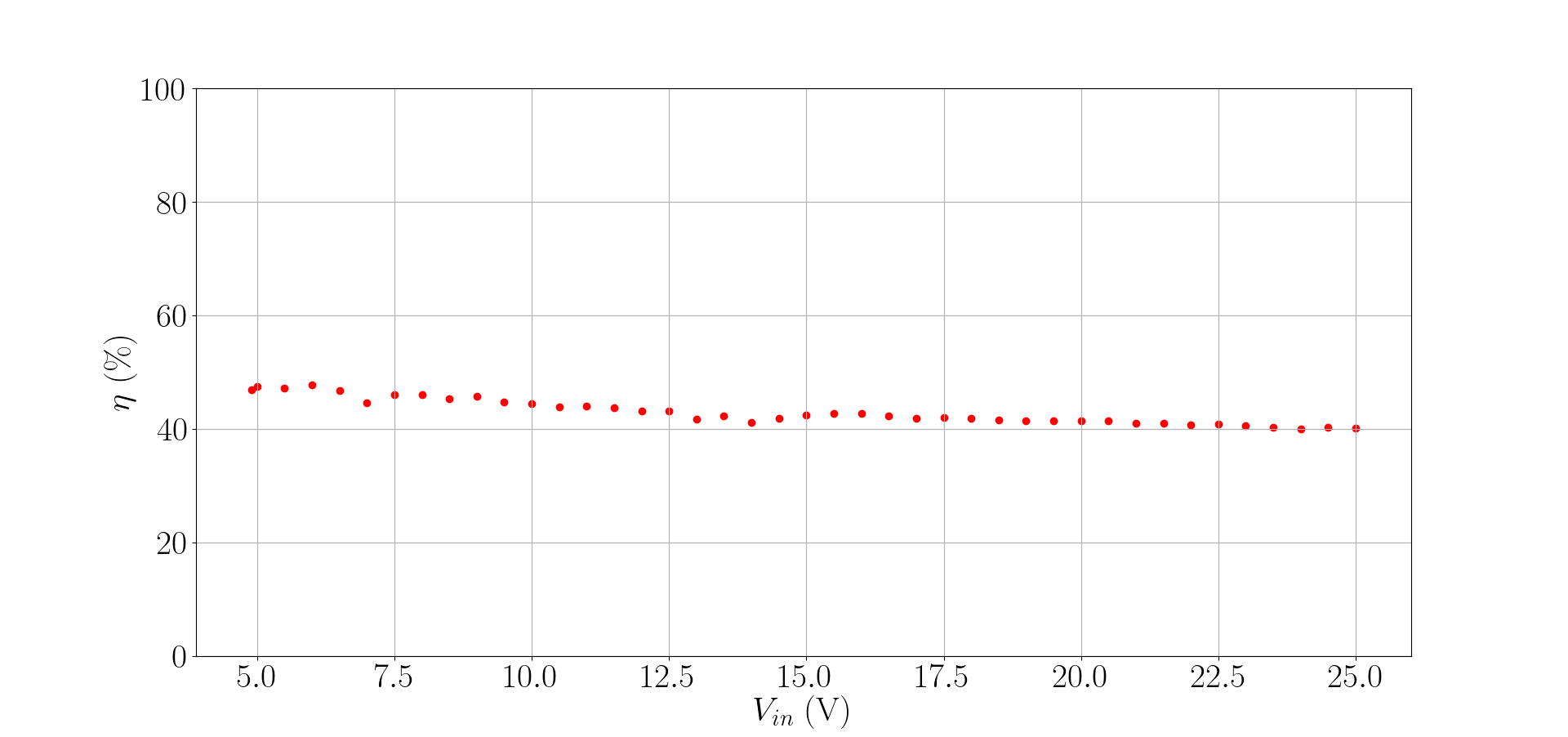}
\caption{
The measured efficiency as a function of $V_{in}$ with the dummy loads is shown.
}
\label{fig:efficiency0T}
\end{figure}

The efficiency of power delivery to the readout electronics is highly related to the chip choice of the DC to DC converter.
The chip in the first stage, TPS53318, has high efficiency of $\geq$ 90\% according to the company  specification \cite{TPS53319}.
However, in the case of linear regulators, the efficiency is lower since there is a large voltage difference between the input and output voltage in linear regulators.
To study the net efficiency of our board, we define the overall efficiency ($\eta$) as
\begin{ceqn}
\begin{equation}
\eta ~ (\%) = \frac{ P_{out} }{ P_{in} } \cdot 100 = \frac{ V_{1} \cdot I_{1} + V_{2} \cdot I_{2} }{ V_{in} \cdot I_{in} }\cdot 100 ,
\label{equation:vFluc}
\end{equation}
\end{ceqn}
where $I_{in}$ and $I_{out}$ are input and output currents of the DC to DC converter.
In order to check the effect from $V_{in}$ to $\eta$,  we measure $\eta$ as a function of $V_{in}$.
Figure \ref{fig:efficiency0T} shows the results of our measurement with the dummy loads under 1 atm.
The overall efficiency is found to be $\geq$ 40 \%, which is a relatively low value for a common DC to DC converter.
This is due to the fact that there are three linear regulators used in our converter for low noise outputs.
Our DC to DC converter can operate from $V_{in}$ = 5.0 V to 25.0 V, which is slightly lager range than the recommended range from company.
There is a tendency that $\eta$ is slowly decreased along with increasing $V_{in}$.
The $P_{in} $ is measured to be 12.4 W, and the power loss $P_{loss} = P_{in} - P_{out}$ is calculated to be 7.4 W as the maximum value when $V_{in}$ is 25.0 V. 
The minimum $P_{loss}$ is 6.4 W when $V_{in}$ is 6.0 V. 

\subsection{Study of heat dissipation} 

\begin{figure}[h!]
   \includegraphics[width=0.6\textwidth]{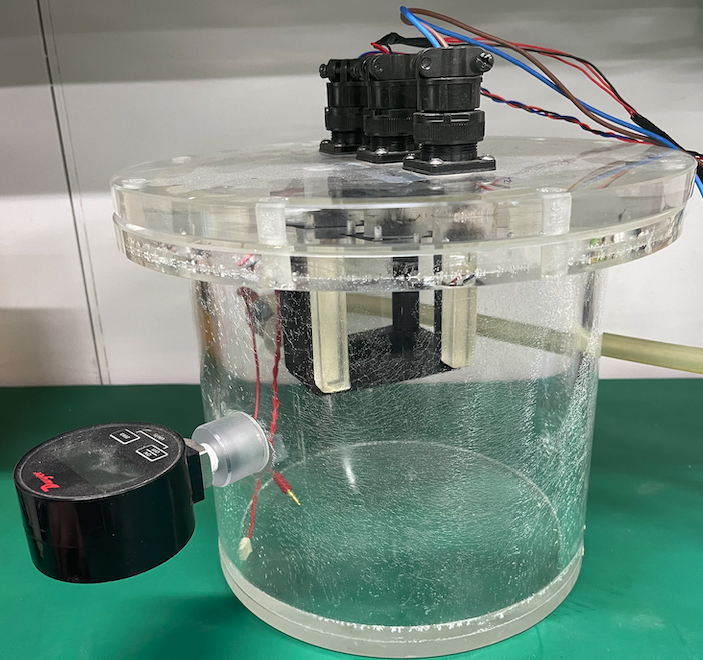}
    \caption{
    The acrylic vacuum chamber for the heat dissipation study is shown. 
    The radius of the chamber is 13 cm, and the height of the chamber is 20 cm with a cylindrical shape.
    Two plastic bars are attached to hold the DC to DC converter in the chamber.
   }
    \label{fig:chamber}
\end{figure}

The DC to DC converter will be placed in a vacuum in the real experiment. 
Therefore, one has to consider the heat dissipation from the converter in that environment.
In order to study this issue, we prepare a small pseudo-vacuum chamber as shown in Fig. \ref{fig:chamber}. 
A small acrylic vacuum chamber with a pump is made to study the heat dissipation from the converter to the outside.
This chamber can set a middle vacuum state ($0.015$ atm), which is similar to the environment of the experiment. 
Under this environment, the temperatures at different locations on the converter are monitored.
The DC to DC converter and temperature sensors are placed in the chamber, and the temperature values are read.
Again, $V_{in}$ is set to be 22 V with the same dummy loads used previously. 

\begin{figure}[h!]
    \includegraphics[width=\textwidth]{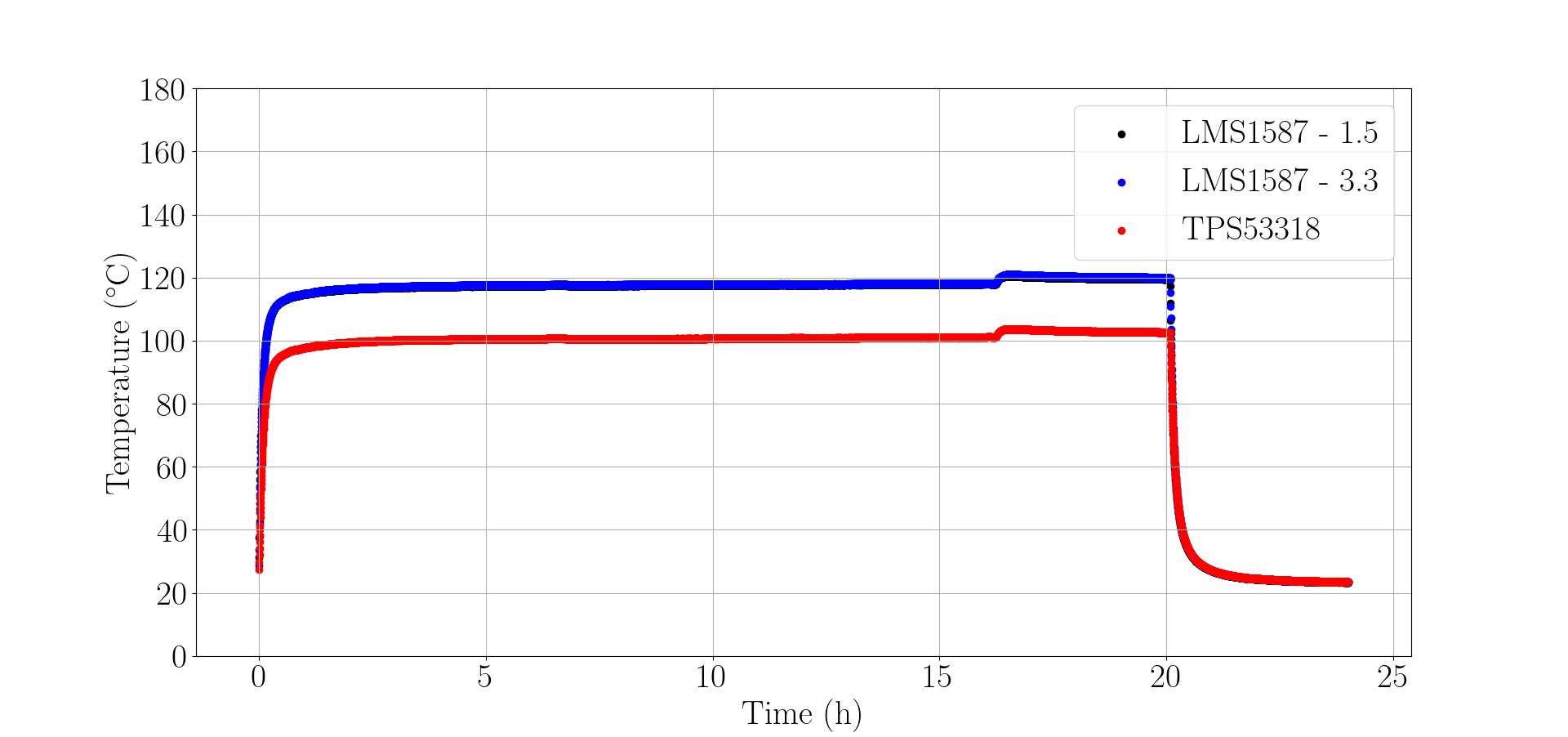}
        \caption{
        The temperature values of the converter as a function of time are shown. 
        The top curve shows the temperature for both linear regulators (overlapped), and the bottom one is for the TPS53318.
        }
\label{fig:tempTest}
\end{figure}

Figure \ref{fig:tempTest} shows that the temperature values of the DC to DC converter in the chamber as a function of time.
The temperature values from TPS53318 and two linear regulators are monitored.
The average temperature is about 100 $^{\circ}$C for TPS53318, and 120 $^{\circ}$C for two linear regulators which is higher than the recommended operation temperature from the company \cite{TPS53319, LMS1585}.
We find that our DC to DC converter works for only about 20 hours in the vacuum chamber.
After 20 hours, the board does not produce output power due to overheating TPS53318 and linear regulators.
During its operation, we find that $P_{out}$ is 5.1 W when, therefore, $P_{loss}$ is 7.3 W.
These results indicate that an active cooling system is required to operate our DC to DC converter in a vacuum.

\section{Conclusions}

We fabricate a DC to DC converter as a low power distribution system, which satisfies special requirements for the muon $g - 2$/EDM experiment at J-PARC.
We simulate the stray magnetic field from the ATI and find that the stray magnetic field at 0.03 ppb.
The voltage stability is found to be less than 3.3 \% at the measurement.
Both stray magnetic field and voltage fluctuation measurement satisfy requirements from the experiment.
Finally, our study of the heat dissipation in vacuum shows that an active cooling system is necessary for the stable operation of the converter in the real experiment.

\section{Acknowledgements}

We would like to sincerely thank Institute for Nuclear and Particle Astrophysics (INPA) at Seoul National University.
This work is supported by a Korea University Grant and the National Research Foundation of Korea Grant funded by the Korean government (Grant No. NRF-2017R1A2B3007018).

\end{document}